\documentclass[aps,prl,twocolumn,showpacs,amsmath,amssymb,superscriptaddress]{revtex4-1}
\usepackage{graphicx}
\usepackage{dcolumn}
\usepackage{amsmath,mathrsfs,amssymb,amsthm}
\usepackage{hhline}
\usepackage{wasysym,enumerate,color}
\usepackage{epstopdf}
\usepackage{verbatim}
\usepackage{hyperref}
\usepackage{color}
\usepackage{ulem} 

\renewcommand{\emph}[1]{\textit{#1}} 

\definecolor{darkgreen}{rgb}{0,0.5,0}
\definecolor{purple}{rgb}{0.35,0,0.35}
\definecolor{orange}{rgb}{1,0.5,0}
\definecolor{darkred}{rgb}{.7,0,0}
\definecolor{darkblue}{rgb}{0,0,.3}
\definecolor{grey}{rgb}{.6,.6,.6}
\definecolor{dimgreen}{rgb}{0.2,0.6,0.1}
\hypersetup{colorlinks,linkcolor=blue,urlcolor=blue,citecolor=blue}

\newcommand{\be}{\begin{equation}}
\newcommand{\ee}{\end{equation}}
\newcommand{\bea}{\begin{eqnarray}}
\newcommand{\eea}{\end{eqnarray}}

\newcommand{\cA}{{\cal A}}
\newcommand{\bx}{{\mathbf x}}
\newcommand{\bs}{{\boldsymbol \sigma}}

\newcommand{\bra}[1]{\langle #1|}
\newcommand{\ket}[1]{|#1\rangle}

\newcommand{\w}{\omega}
\newcommand{\s}{\sigma}

\def\erf{\eqref}
\newcommand{\vev}[1]{\left\langle #1 \right\rangle}

\newcommand{\ud}          {\mathrm d}

\newcommand\fii           {\varphi}

\newcommand\p             {\partial}
\newcommand\kB             {k_\text{B}}

\renewcommand\th          {\theta}

\begin{document}
\title{Hybrid semiclassical theory of quantum quenches in one dimensional systems}
\author{C\u at\u alin Pa\c scu Moca}
\affiliation{BME-MTA Exotic Quantum Phase Group, Institute of Physics, Budapest University of Technology and Economics,
H-1111 Budapest, Hungary}
\affiliation{Department of Physics, University of Oradea, 410087, Oradea, Romania}
\author{M\'arton Kormos}
\affiliation{BME-MTA Statistical Field Theory Research Group, Institute of Physics, Budapest University of Technology and Economics,
H-1111 Budapest, Hungary}
\author{Gergely Zar\' and}
\affiliation{BME-MTA Exotic Quantum Phase Group, Institute of Physics, Budapest University of Technology and Economics, H-1111 Budapest, Hungary}

\frenchspacing

\date{\today}
\begin{abstract}
We develop a hybrid semiclassical method to study the  time evolution  of one dimensional quantum  systems in and out of equilibrium. Our method handles internal degrees of freedom completely quantum mechanically by a modified time evolving block decimation method, while treating orbital quasiparticle motion classically.  We can follow  dynamics up to timescales well beyond the reach of standard numerical methods to observe the crossover between pre-equilibrated and locally phase equilibrated states.
 As an application, we investigate the quench dynamics and phase fluctuations of a pair of tunnel coupled one dimensional Bose condensates. 
 We demonstrate the emergence of soliton-collision induced phase propagation, soliton-entropy production and multistep thermalization. 
 Our method can be applied to a wide range of gapped one-dimensional systems.
\end{abstract}

\maketitle

Fundamental questions concerning the coherent time evolution, relaxation and thermalization of isolated quantum systems have been brought into the focus of attention by  recent progress in experimental techniques \cite{Polkovnikov2011,Kinoshita2006, Hofferberth2007, Gring2012}. Experiments on cold atomic gases allow us to engineer  a broad range of lattice and continuum Hamiltonians in a controlled fashion, and to monitor the coherent time evolution of these systems through measuring multi-point correlation functions \cite{Schweigler2015}, accessing the quantum state via site-resolved quantum microscopy \cite{Bakr2009}, and even measuring the entanglement properties of the system~\cite{Islam2015}. These experiments as well as
ongoing matter wave interferometry~\cite{schumm2005} experiments
call for the development of new  analytical and numerical methods that are able to describe  non-equilibrium dynamics in closed interacting quantum systems, 
and address fundamental questions of non-equilibrium thermodynamics, such as thermalization, entropy production, 
or the fate of pre-thermalized states.

\begin{figure}[!b]
\includegraphics[width=0.75\columnwidth]{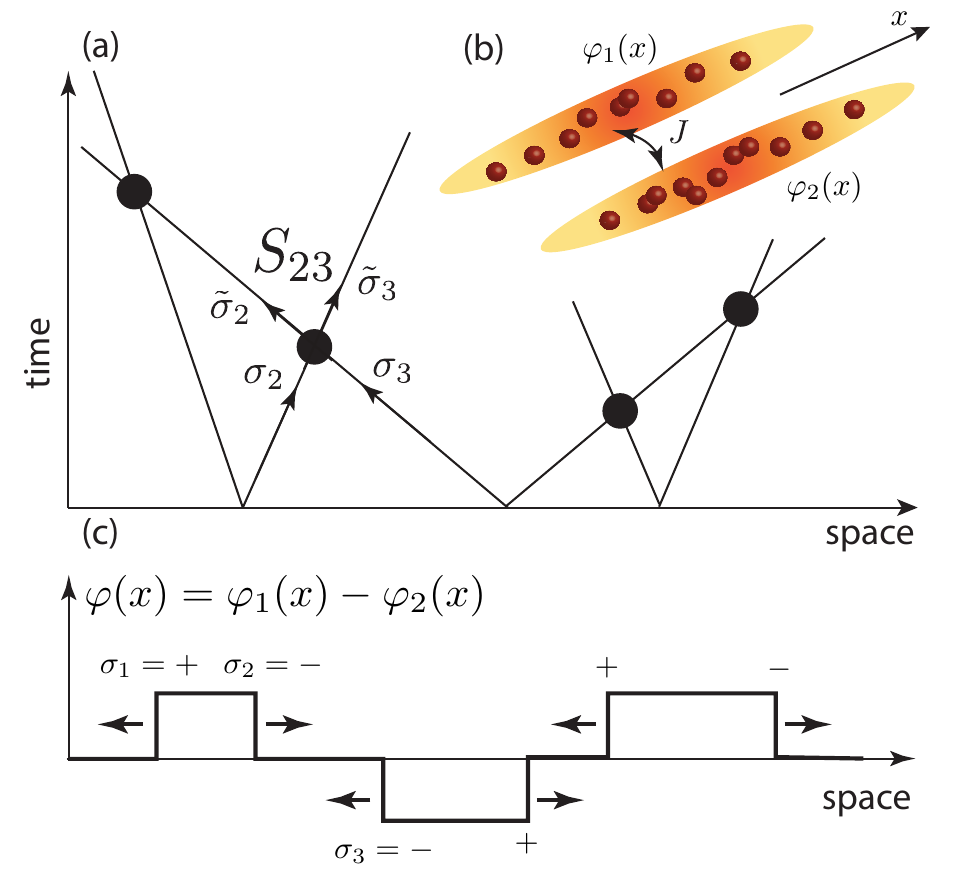}
\caption{(Color online) (a) Space-time diagram of 
quasiparticle trajectories. Labels $\sigma=\pm$ represent 
internal soliton/antisoliton quantum numbers. 
Collisions  are described by the quantum mechanical $S$-matrix (solid black dot).  
(b) One dimensional Bose condensates coupled by quantum tunneling.
Dynamics
of the relative phase 
is described by the sine--Gordon model.
(c) The quench creates  soliton-antisoliton  
pairs   at $t=0$. 
Arrows indicate  directions of  propagation. }
\label{fig:traj-cloud}
\end{figure}

Here we propose a novel method we dub ``semi-semiclassical'' (sSC), 
able to reach time scales much beyond conventional methods~\cite{White2004, Schollwock:2011,Vidal2003} and to capture 
the fundamental phenomena of pre-thermalization \cite{Berges2004, Buchhold2015} as well as local equilibration in great detail. 
Our method  hybridizes a semiclassical (SC) approach with time-evolving block decimation (TEBD): we compute  the 
time evolution of the internal degrees of freedom completely quantum mechanically using TEBD,
while treating the orbital motion semiclassically.  
 As a proof of principle, we use our method  to describe quantum quenches in the 
sine-Gordon model, relevant for  two coupled 1D quasicondensates (see Fig.~\ref{fig:traj-cloud}.b), and  studied intensely both theoretically \cite{Gritsev2007,Iucci2010,Bertini2014,inflation,Delfino2016} and in matter wave interference experiments using nanofabricated 
 atom chips \cite{Hofferberth2007, Kuhnert2013}.
 Our sSC method is demonstrated to capture  multistep thermalization and soliton-collision induced entropy production efficiently,
and is applicable to one dimensional gapped systems having stable quasiparticle excitations
~\footnote{The most common methods, TEBD~\cite{Vidal2003} and 
t-DMRG~\cite{White2004, Schollwock:2011}, are   both 
limited by finite size effects, and typically work only for short times, preceding thermalization. Quantum Monte Carlo techniques may be a promising direction (see G. Carleo et al., (2016) arXiv:1612.06392).}.

Our method is based on a semiclassical (SC) approach, originally developed to study the dynamics of  finite temperature systems in equilibrium~\cite{Sachdev1996,Damle2005,Rapp2006} and later successfully applied to quantum quenches~\cite{Rieger2011,Rossini2009,Evangelisti2013,Kormos2015}, i.e. to non-equilibrium situations in which the system evolves unitarily starting from some prepared initial state. 
The SC method is applicable in gapped systems whenever the quasiparticles' Compton wavelength (or the thermal wavelength) is much shorter than their average separation, $d= \rho^{-1},$ with $\rho$ the quasiparticle density.
If the energy of the initial state is not too high, it acts as a weak source of almost pointlike quasiparticles,  
following classical trajectories~\cite{CalabreseCardy2006,calabrese2005evolution} (see Fig. \ref{fig:traj-cloud}.a). 
 In the standard SC treatment,  quasiparticles are furthermore 
assumed to move slowly and thus  collisions are described by a universal purely reflective  scattering matrix~\cite{Sachdev1997}. 
This universal SC (uSC)  approach 
permits the derivation of precious analytical results~\cite{Damle2005,Rapp2006,Evangelisti2013,Kormos2015}, but also 
suffers from artifacts~\cite{Rapp2006,Kormos2015};  certain correlation functions and expectation values fail to  decay 
and internal degrees of freedom remain just locally entangled.
Our method  eliminates all these shortcomings  
\footnote{
We follow the  nomenclature  of Ref. \cite{Sachdev1996}. There  low density implies small quasiparticle momenta, $p$, and since $p\sim \hbar,$  
the limits $p\to 0$ and  $\hbar\to0$  are equivalent,  explaining the term `semiclassical'.  We lift the restriction to small $p$ but require low quasiparticle density.}.

\paragraph{Model.---} The sine-Gordon model is defined as 
\begin{multline}
H=\frac{\hbar c}2 \int\ud x \left[\frac{2\pi}{K}  \Pi(x)^2+\frac{K}{2\pi} [\p_x\fii(x)]^2\right]-\\
 \int\ud x\,2\Delta_0\cos[\fii(x)]\,,
\end{multline}
with $\Pi(x)$ denoting the field conjugate to $\varphi(x)$, and $c$  the speed of sound. 
In case of  coupled condensates,  $\varphi(x)=\varphi_1(x)-\varphi_2(x)$ represents the relative phase of the condensates,
$K$   stands for the Luttinger parameter
of each condensate, and  the cosine term accounts for  Josephson tunneling between them \cite{Gritsev2007a}.    
For locally interacting bosons $K \ge 1$, with $K\gg1$ corresponding to weak interactions. 
The cosine perturbation is relevant for $K>1/4,$ and 
for $K<1/2$ there are only kink excitations in the system, solitons and antisolitons. In the classical field theory, 
these kinks interpolate between neighboring minima of the cosine potential and have topological charges $\sigma = \pm$ 
corresponding to changes $\varphi \to \varphi\pm 2\pi$, respectively.  
For $K>1/2$ their bound states, the so-called breathers  are also present.   
However, having no topological charge, breathers are supposed to be irrelevant for the long time (large distance)  phase correlations studied 
here, and shall therefore be neglected in what follows.

We illustrate our method on the out of equilibrium time evolution of the coupled 1D condensates after changing the potential 
barrier that separates them. For small  changes in the model parameters the quench is perturbative, and the gas of quasiparticles (kinks) created in the quench will be dilute such that the SC method can be applied~\cite{Rieger2011,Werner17}. 
Furthermore, by momentum conservation,  a homogeneous but spatially localized perturbation gives rise to pairs of kinks flying away from each other with the same velocity and opposite topological charge.  
Thus 
the post-quench state will be populated by independent soliton-antisoliton pairs created with a velocity distribution $f(v)$.  The precise form of $f(v)$ depends on  details of the quench protocol, but  turns out to be unimportant in the present calculation. The initial state considered here resembles to the ones appearing 
in previous studies of split condensates \cite{Kitagawa2011,Foini2014} as well as other systems \cite{Silva2008,Sotiriadis2010a,Fioretto2010,deNardis2014,Pozsgayoverlaps2013,Bertini2014,Horvath2016}, where the initial state was taken to be a coherent superposition of uncorrelated zero momentum quasiparticle pairs. 
Within our sSC approach, however,  phase coherence of the pairs does not play any role. 

\paragraph{The semi-semiclassical (sSC) method.---}
In the  SC approach, the quantum mechanical average is replaced by an ensemble average over 
initial positions and velocities of solition-antisoliton pairs. The corresponding 
semiclassical configurations consist of space-time diagrams with pairs of straight lines (kink trajectories) originating from the $t=0$ axis, distributed independently from each other, uniformly in space with random slopes corresponding to 
 the distribution $f(v)$ (see  Fig. \ref{fig:traj-cloud}.a),  and an initial distribution of paired charges.

Since quantum mechanical effects become relevant only at collisions, we can approximately factorize  the many body wave function into an orbital and a charge  part of the form 
$\ket {\Psi(\bx, \bs, t)} \approx \mathcal{S} \ket {\Psi_{\rm orb}(\bx, t)}\otimes \ket {\chi( \bs, t)},$ where
\begin{eqnarray}
\ket {\Psi_{\rm orb}(\bx, t)}&=&
\int {\rm d}^Nx
\prod_{j=1}^{N}\delta(x_{j}-x^0_j-v_{j}t)\ket{x_1,\dots x_N}\,,\label{orb}
\nonumber
\\
\ket {\chi( \bs, t)}&=&\sum_{\s_{1},\s_{2},\dots}\cA_{\s_{1}\s_{2}\dots\s_{N}}\ket{\s_{1}\s_{2}\dots\s_{N}}\label{charge}\,.
\end{eqnarray}
Here $x^0_j$ denotes the 
initial position of the $j^{\rm th}$ kink of velocity $v_j$ and topological charge $\sigma_j,$ and $\mathcal{S}$ stands for symmetrization.

In the original uSC approach  charges are treated classically: quasiparticles  scatter as impenetrable
billiard balls while  preserving their charges. This follows from the perfectly reflective 
universal low energy two-body $S$-matrix,
$S_{\s_{1}\s_{2}}^{\tilde \s_{1}\tilde \s_{2}} = (-1)\delta_{\s_{1}\tilde \s_{1}}
\delta_{\s_{2}\tilde\s_{2}}$ (notice the labeling convention), describing the scattering of 
quasiparticles with vanishing momenta.
Using this perfectly reflective S-matrix allows one to obtain a number of exact results for thermal gases 
and near adiabatic quantum quenches~\cite{Damle2005,Rapp2006,Evangelisti2013,Kormos2015}. 
 However, as soon as the quench is not slow enough, faster quasiparticles are inevitably created. Collisions involving these faster particles are not captured by the universal scattering matrix, and the true velocity dependent scattering matrix must be used (cf. Fig. \ref{fig:traj-cloud}.a). The sine--Gordon model is integrable and this two-particle $S$-matrix is exactly known \cite{Zamolodchikov1979,SM}. 
The matrix elements 
$S_{+-}^{+-}= S_\text{R}(v_1,v_2)$ and
$S_{+-}^{-+}= S_\text{T}(v_1,v_2)$, in particular, describe reflection and transmission, respectively. 
They satisfy $|S_\text{T}|^2+|S_\text{R}|^2=1,$ and for small velocities transmission vanishes as  $|S_\text{T}(v_1,v_2)|^2\propto(v_1-v_2)^2.$

We can incorporate the non-trivial $S$-matrix in two different ways. In the first method,  charges are still treated classically, but at each collision either a perfect transmission or perfect reflection takes place with probabilities given by the modulus square of the $S$-matrix elements. This method neglects interference effects but can be implemented as a classical Monte--Carlo simulation. A more refined approach is to treat the charge part of the wave function in Eq. \erf{charge} in a fully quantum mechanical manner with an MPS-based method. As the quench protocol generates neutral kink pairs with equal probability amplitude,  the initial  wave function is a dimerized state, i.e. a product of  Bell pairs: 
\be
\label{chi0}
\ket {\chi(\bs, t=0)} = \prod_{j=1}^{N/2}\frac{\ket+_{2j-1}\ket-_{2j} + \ket-_{2j-1}\ket+_{2j}}{\sqrt{2}}\,.
\ee
The charge wave function evolves only through collisions and it is frozen between collisions. If line $j$ and $j+1$ intersect at time $t_{k}$, the change of the wave function is 
\be
\ket {\chi(\bs, t_{k,+})} = \hat S_{j,j+1}(t_{k})\ket {\chi(\bs, t_{k,-})}\,,
\ee
where $\hat S_{j,j+1}(t_k)$ acts nontrivially only on charges $j$ and $j+1$ via the 2-body $S$-matrix evaluated at the velocities of the colliding quasiparticles.
In this way we mapped the dynamics of the charges to that of an effective quantum spin chain. The time evolution operator of this spin chain depends on the underlying semiclassical trajectories and is given as a product of local unitary two-body operators, efficiently treated by an MPS-based  algorithm.

\paragraph{Phase distribution.---}
The space 
resolved relative phase $\varphi(x)$ 
of the condensates can be measured directly
via interferometry experiments  \cite{Hofferberth2007,Jo2007,Kuhnert2013,Schweigler2015}. 
We first use the classical Monte Carlo approach to determine the full phase distribution after the quench,
 a quantity also analyzed experimentally~ \cite{Schweigler2015}. 
 in the experiment of   Ref.~\cite{Schweigler2015}. 
Semiclassically, each kink 
is a domain wall that separates two domains with a phase difference of  $\pm2\pi.$ 
Domains separated by $s$ kinks have a phase difference $\Delta\varphi= 
 2\pi\sum_{i=1}^{s} \sigma_i$.  
The phase in a given domain being constant, its  
distribution function is  a sum of weighted Dirac-delta peaks 
located at integer multiples of $2\pi$, 
\be
\label{eq:cn}
P(\varphi, t) = \sum_{n\,\in Z}c_{\,2\pi n}(t)\,\delta\bigl (\fii-2\pi n\bigr)\,. 
\ee
In  experiments, these delta peaks get broadened by quantum and thermal fluctuations. To take this into account, we estimated the phase fluctuations $\langle \fii(x)^2 \rangle$ around the minima of the cosine potential for typical 
parameters
\cite{SM}, and broadened the delta-functions accordingly.
\begin{figure}[!tb]
\includegraphics[width=0.84\columnwidth]{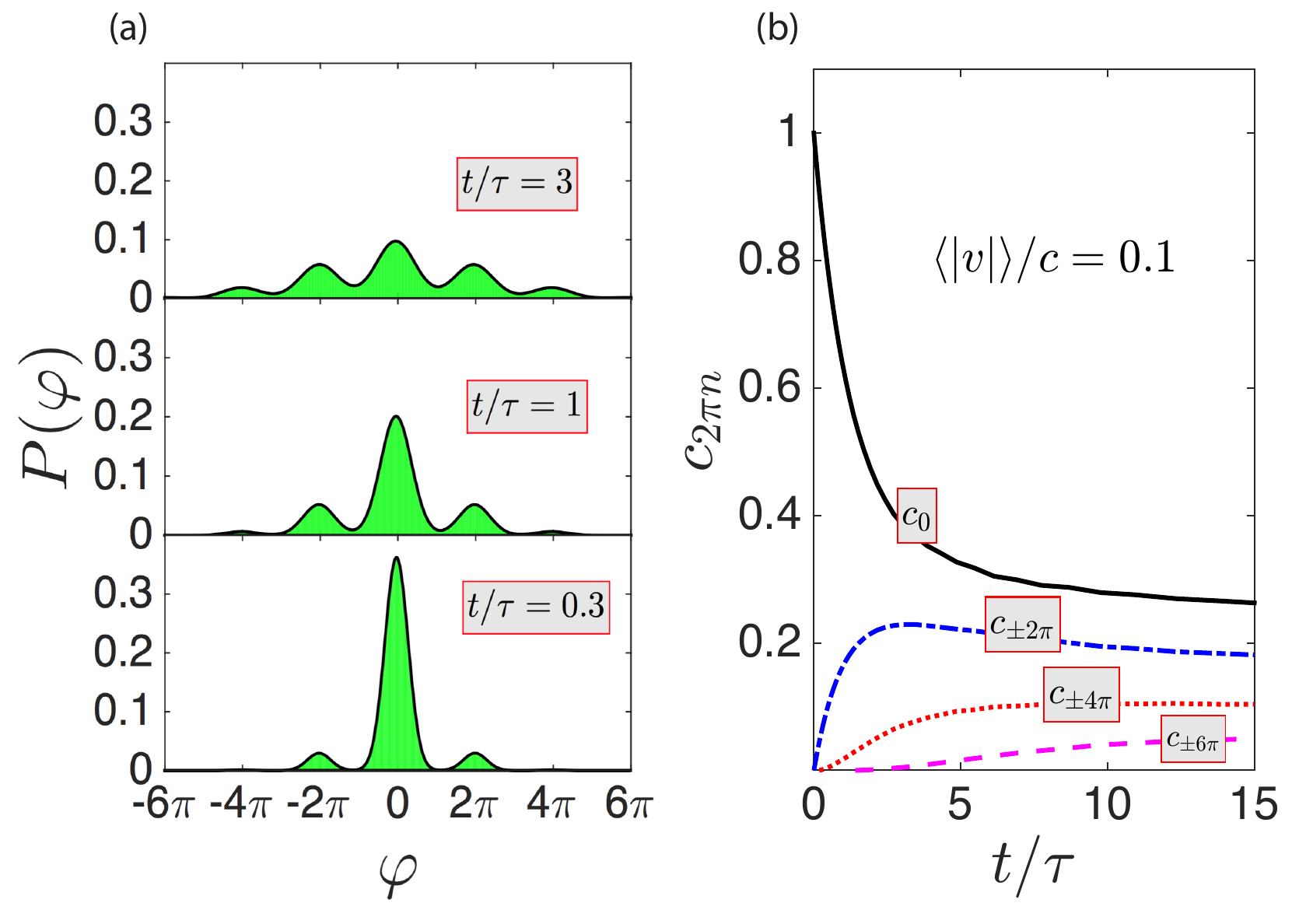}
\caption{(a) Phase distributions  $P(\varphi,t)$ at different times for a system of $N = 24$ kinks with 
$\langle |v|\rangle = 0.1\, c$ and Luttinger parameter $K=9.82$. (b) Time dependence of the weights $c_{2\pi n}$ defined in  Eq.~\eqref{eq:cn}. }
\label{fig:phase_distribution}
\end{figure}

The resulting phase distribution   $P(\varphi, t)$ is shown in Fig.~\ref{fig:phase_distribution}.a 
 at different instances for a typical small quench.  
For the momentum distribution of quasiparticles, $f(p)$,  we used the ansatz $f(p) \propto p^2 \exp(-p^2/ p_0^2)$,
motivated by overlap expressions in the transverse field Ising model~\cite{CEF1}.
 Immediately after the quench, there is a single central peak 
at $\varphi=0$ and $c_0(0)=1$, but side peaks emerge as the system evolves in time.
The two side peaks at $\fii=\pm 2\pi$ emerge already at the uSC level:  
initially,  domains of phase  $\fii=\pm2\pi$ grow ballistically between separating soliton-antisoliton pairs (see Fig.~1.c), 
therefore $c_0$ decreases while $c_{\pm 2\pi}$ increase  linearly in time.  Within the uSC approach,  
the  weights  $c_{\pm 2\pi}$ 
saturate at   
$c_{\pm 2\pi}\to1/4$ once the coordinates of the kinks become randomized, while   the central 
peak levels off at  $c_0\to 1/2$.  We observe this '\emph{pre-thermalized}' behavior by sSC at intermediate times  in simulations with small quasiparticle velocities  $\langle |v|\rangle /c \lesssim 0.05$. 
However, after a fast relaxation,   collisions start to dominate beyond the collision time, 
$t\gtrsim\tau = \rho \int_0^\infty \ud v\, vf(v),$
 and  \emph{phase propagation} takes place due to transmissive collisions, 
giving rise  
 to  domains of phases  $\fii=\pm4\pi,\pm6\pi,\dots$,  
  and the emergence  of  further side peaks 
 (see Fig.~\ref{fig:phase_distribution}.b).
 These are  all expected  to vanish  as $c_{2\pi n }\sim 1/\sqrt{t}$ for very long times,  reflecting  the expected diffusive nature of  phase propagation, as indeed supported by our numerical results (see \cite{SM}).

\begin{figure}[!tb]
\includegraphics[width=0.74\columnwidth]{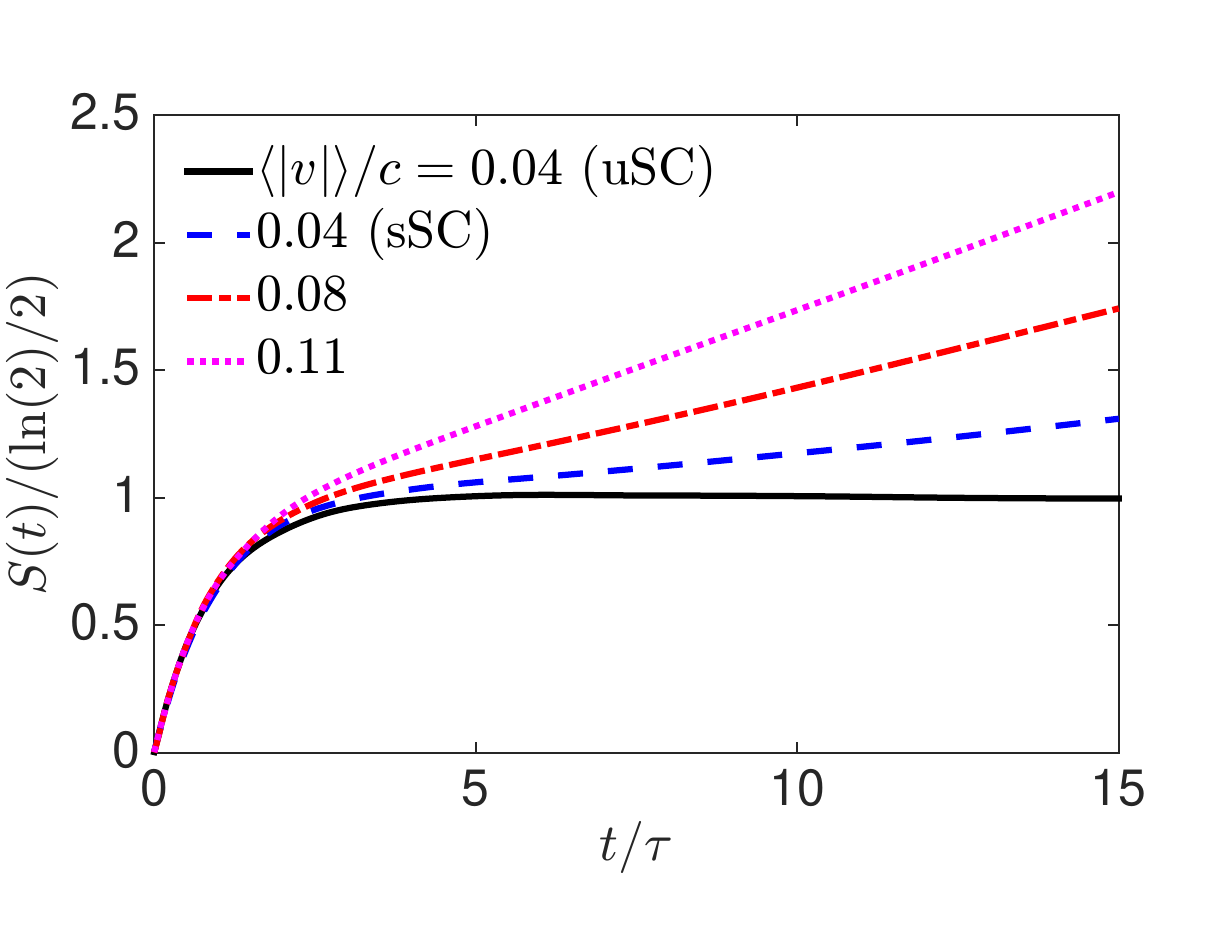}
\caption{ Dynamics of  charge entanglement entropy after the quench for 
the same parameters as in Fig.~\ref{fig:phase_distribution}. The solid
black line denotes the uSC result with a fully reflective  $S$-matrix. 
}
\label{fig:entropy}
\end{figure}

\paragraph{Entanglement entropy.---}
Our method is able to follow the propagation and growth of entanglement in the  
charge sector. The most widely used entanglement measure is the entanglement entropy of a subsystem $A$, defined as the von Neumann entropy of its reduced density matrix: $S = -\rm Tr_A \big [  \rho_{A}\log \rho_{A} \big ]$ where $\rho_A=\mathrm{Tr}_{\bar A}  \ket{{\Psi(\bx, \bs, t)}}\bra{{\Psi(\bx, \bs, t)}}.$ Here we focus on  the simple case when  $A$ is  half of the total system.
The initial charge wave function  $\ket {\chi}$ 
has alternating bond entanglement entropies 
of $\log 2$ and 0. At $t=0^{+}$, the spatial extension of the entangled bonds is zero, so with probability 1 the cut between the two halves of the system falls between entangled pairs resulting in zero entanglement entropy. 
Within the uSC approximation, at $t>0$, pairs with one member on the left and another member on the right entangle the two halves of the system.  Since the uSC $S$-matrix is fully reflective, each kink remains maximally entangled with its original partner, but entanglement remains local. 
At very long times, the cut between the two halves of the system either falls between pairs or cuts a pair in two with probability 1/2. Consequently, the entanglement entropy saturates at a value $S_\text{sat}=\log 2/2.$ 
The  full time evolution of the entropy is simple to calculate within the 
uSC approach  which yields  an exponential relaxation   to  $S_\text{sat}$, 
$S_{\rm uSC}(t) = (1-e^{-t/\tau})\ln 2 /2,$ (solid black line in Fig.~\ref{fig:entropy}). 
The saturation of $S_{\rm uSC}(t)$ is clearly an artefact of the uSC method. In reality, the topological charges  of remote quasiparticles become entangled with time due to transmissive scattering processes. 
The sSC approach captures the corresponding entropy production: the
 entanglement entropy does not saturate after the initial transient, but is found to grow \emph{ linearly} in time without bound in an infinite system, $S(t)=\alpha\,t$, with  a growth rate $\alpha\propto \langle |v|\rangle$. 

\begin{figure}[!tb]
\includegraphics[width=0.76\columnwidth]{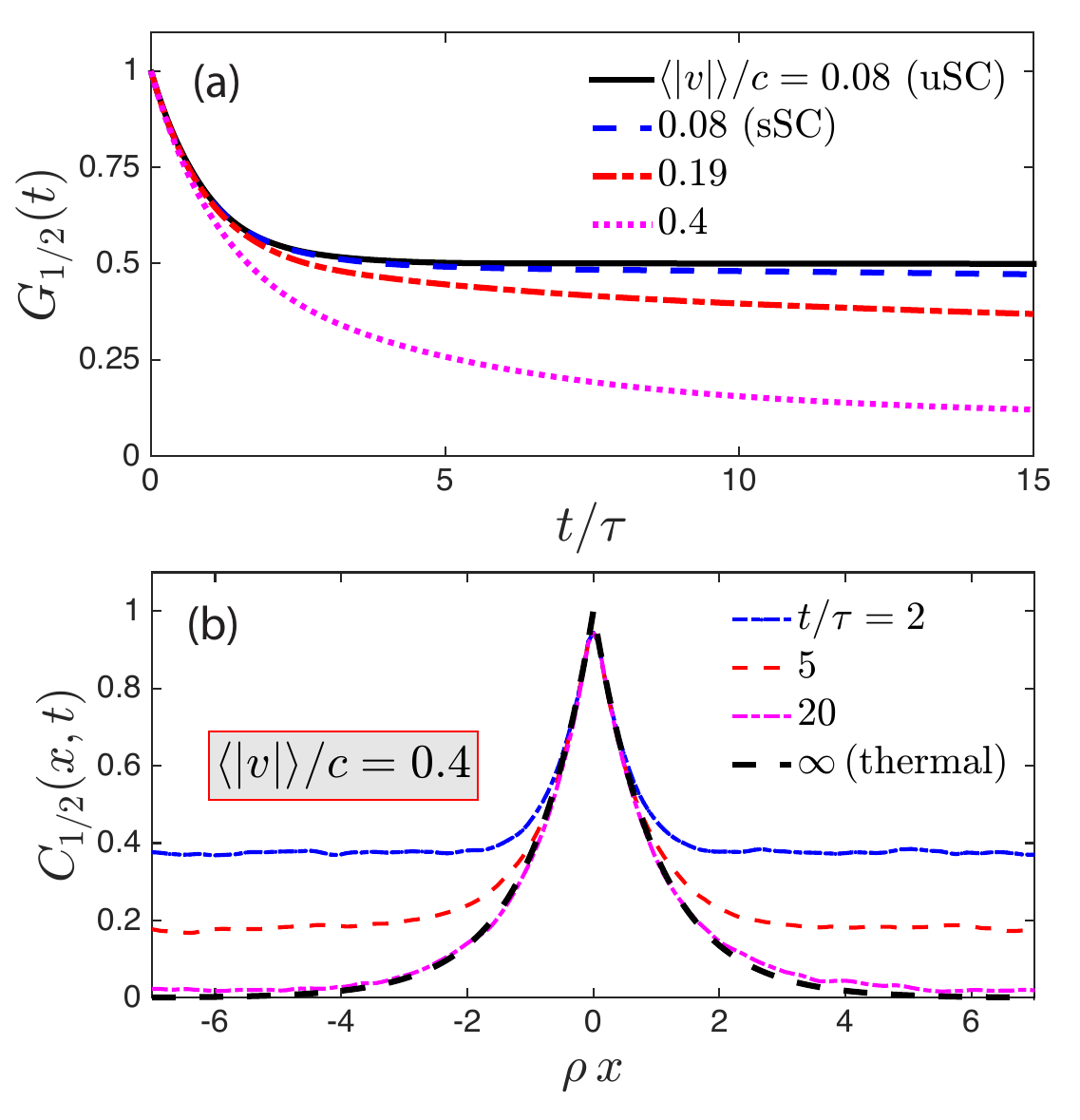}
\caption{(a) Relaxation of the expectation value $\langle  e^{i \varphi(x,t)/2}\rangle$
for different average velocities.
The solid black line represents the 
uSC result (fully reflective S-matrix).
(b) Equal time correlation function $\langle  e^{i \varphi(x,t)/2} e^{-i \varphi(x', t)/2}
\rangle$ at different times calculated with the sSC method.
The black dashed line is the thermal equilibrium result in the uSC approximation.
}
\label{fig:correlators}
\end{figure}

\paragraph{Correlation functions and equilibration.---}
The sSC method is also suitable for computing the out of equilibrium evolution of correlation functions.   We consider first the expectation values $G_{\alpha}(t) = \langle  e^{i\alpha \varphi(x,t)}\rangle
=  \langle  e^{i\alpha \varphi(x,t)}e^{-i\alpha \varphi(x,0)}\rangle$. The function $G_1(t)$  is essentially the 
coherence factor,  directly accessible through matter wave interferometry~\cite{Hofferberth2007}.
The standard uSC approach has been recently used to compute  $G_\alpha(t)$~\cite{Kormos2015}, 
 which  was found to decay exponentially to a  value $\cos^2(\pi \alpha)$, apart from a  prefactor 
   incorporating 
vacuum fluctuations, now set to one (see  Fig.~\ref{fig:correlators}.a). This surprising behavior is 
related to the fact that in the uSC approach the phase is pinned to the values 
$\fii = 0, \pm 2\pi$. In contrast, within the sSC approach,  
the phase  meanders with time and   $G_{\alpha}(t\to \infty )$ no longer remains finite; 
after a fast  transient described approximately by  uSC,   $G_{\alpha}(t)$ is found to decay  
exponentially to zero  with a rate  depending on $\langle|v| \rangle/c $. 
The equal time two-point correlation functions,
\begin{equation}
C_{\alpha}(x-x'; t) = \langle  e^{i\alpha \varphi(x,t)} e^{-i \alpha\varphi(x', t)}\,,
\rangle \,
\label{eq:correlator}
\end{equation}
are also accessible experimentally~\cite{Langen2013}, 
and the  integrated quantity, $\int\ud x\,\ud x' C_{1}(x-x';t),$ is directly related to the contrast of the interference fringes.
Fig.~\ref{fig:correlators}.b shows the time evolution of $C_{\alpha}(x;t)$ for $\alpha= 1/2$.
Here again, the uSC approach yields a quick relaxation to a state, where the phase is pinned and, accordingly, long-ranged correlations persist,  $C_{\alpha}(|x|,t\to \infty) \sim \cos^4(\pi \alpha)$.

In sharp contrast to   traditional uSC, within the sSC we find that 
$C_\alpha(x-x', \infty)= 
\exp(-2\sin^{2}(\pi\alpha)\rho|x-x'|)$  which formally coincides with
the \emph{thermal} equilibrium correlator  
computed within the standard finite temperature SC approach \cite{Damle2005}. 
In our case, however, the final state is \emph{not} thermal:  in the spirit of Generalized Gibbs Ensemble (GGE) the density $\rho$ is conserved and set by the initial state rather then a single effective temperature~\cite{rigol2007}.
The final state we obtain should rather be viewed as a 
\emph{pre-thermalized} state, where local phase correlations and expectation 
values of vertex operators look thermal, but the velocity distribution of the quasiparticles
remains non-thermal, and the quasiparticles' average energy is not related to their density.
The sSC  method thus apparently captures aspects of local phase equilibration  and  ``pre-thermalization''.

Equilibration is thus predicted to take place in several steps in a split condensate. For small values of $\langle |v|\rangle /c$, first a quick relaxation occurs to a first pre-equilibrated state
with pinned phase and a non-thermal quasiparticle velocity distribution, but with kink 
positions randomized. Next, the coherent evolution of the 
charge wave function gives rise to phase (quantum)  propagation. At this stage, phase correlation functions relax to their GGE values. To reach a truly thermal state with thermal quasiparticle velocity distribution  a further relaxation step and coupling to some 
external environment such as the symmetrical phase mode 
are ultimately needed in the coupled condensate experiment.

\paragraph{Conclusions.---}  

The versatile semi-semiclassical method developed here has a broad range of applicability. It is suitable for studying the dynamics of 
lattice systems and  spin chains \cite{Werner17}  as well as continuum 1D systems. The differences between these  translate into differences in the nature of quasiparticles and  their 2-body S-matrix, which makes our method ideal for identifying universal aspects of the dynamics. The semiclassical aspect allows for heuristic interpretations and relatively long simulation times, while the quantum description of the scattering of quasiparticles allows us to go beyond the limitations of the standard uSC approach and  opens the way to study the propagation of entanglement. Possible connections with the recent work \cite{Alba2016} would be interesting to analyze.
Although here we focused on the non-equilibrium time evolution, the method can also be used to investigate dynamical correlation functions at finite temperature beyond the uSC approximation, and is well-suited to study non-equilibrium dynamics in inhomogeneous 1D
systems~\cite{Kormos17}.

\paragraph{Acknowledgements.} 
We  thank J\" org Schmiedmayer, Eugene Demler and Ehud Altman for fruitful discussions.
This work was supported by the Hungarian research fund OTKA under grant Nos. K105149 and SNN118028. 
M.K. was partially supported by a J\'anos Bolyai  Scholarship and a Pr\'emium Postdoctoral Fellowship of the HAS.

\bibliographystyle{apsrev4-1}

\bibliography{semi-semiclassic_11g}

\onecolumngrid
\appendix

\renewcommand{\thesection}{S-\Roman{section}}
\renewcommand{\theequation}{S-\arabic{equation}}
\renewcommand{\thefigure}{S-\arabic{figure}}

\bigskip
\bigskip

\centerline{\bf \Large Supplemental Material}

\bigskip
\bigskip

\section{The Hamiltonian and sine--Gordon description of coupled quasicondensates}\label{app1}

The Hamiltonian of two coupled 1D condensates is given by~\cite{Gritsev2007a}

\begin{multline}
H=
\sum_{j=1,2}\int\ud x \left\{\frac{\hbar^2}{2m}\p_x\psi_j^\dag(x)\p_x\psi_j(x)+\frac{g}2\psi_j^\dag(x)\psi_j^\dag(x)\psi_j(x)\psi_j(x)+[V(x)-\mu]\psi_j^\dag(x)\psi_j(x)\right\}\\
-\hbar J\int\ud x \left[\psi_1^\dag(x)\psi_2(x)+\psi_2^\dag(x)\psi_1(x)\right]\,,
\end{multline}
where $\psi_1(x),\psi_2(x)$ are the bosonic fields in the two condensates, $V(x)$ is the longitudinal potential which we neglect in the following, $J$ is the tunnel coupling between the condensates, and 
\be
g=\frac{2\hbar^2a_s}{ml_\perp^2}\left(1-1.0325\frac{a_s}{l_\perp}\right)^{-1}
\ee
is the strength of the atom-atom interaction within each condensate. Here $l_\perp=\sqrt{\hbar/(m\w_\perp)}$ with $\w_\perp$ being the frequency of the radial confinining potential and $a_s$ denotes the $s$-wave scattering length of the atoms. 
The strength of interaction is often parameterized by the dimensionless combination
\be
\gamma=\frac{m g}{\hbar^2n}\,.
\ee
In the absence of coupling, the condensates can be described within the bosonization framework by the Hamiltonians
\be
H_j=\frac{\hbar c}2 \int\ud x \left\{\frac\pi{K}\, \Pi_j(x)^2+\frac{K}\pi [\p_x\fii_j(x)]^2\right\}\,,
\ee
where $[\fii_j(x),\Pi_j(x')]=i\delta(x-x').$  The speed of sound, $c,$ and the Luttinger parameter $K$ can be computed from the exact Bethe Ansatz solution of the model. The asymptotic expansions for small and large couplings are
\begin{subequations}
\begin{align}
K&\approx\frac\pi{\sqrt\gamma}\left(1-\frac{\sqrt\gamma}{2\pi}\right)^{-1/2} \approx\hbar\pi\sqrt{\frac{n}{mg}}\,,&
c&\approx\sqrt{\frac{ng}m}&  \text{for } &\gamma\lesssim10\,,\\
K&\approx(1+4/\gamma)\,,& c&\approx\hbar\pi n/m& \text{for }&\gamma\gg1\,.
\end{align}
\end{subequations}
Due to Galilean invariance, $cK=\hbar n\pi/m$ holds for all $\gamma.$ 
The density fluctuations are suppressed at wavelengths smaller than the healing length which is also used as a short distance cutoff. For small $\gamma$ it is
\be
\xi_\text{h}=1/(n\sqrt{\gamma})=\hbar/\sqrt{mgn} \approx \hbar/mc\,,
\ee
while at strong coupling $\xi_\text{h}\approx1/n.$ 
The coupling between the condensates is captured by the term $2\Delta\cos(\fii_1-\fii_2),$ where $\Delta\approx\hbar Jn$ for weak coupling, but it can be renormalized at strong interactions.
It is convenient to introduce the total and relative phase $\fii_\pm=\fii_1\pm\fii_2$ and  $\Pi_\pm=(\Pi_1\pm\Pi_2)/2$ in terms of which the two Hamiltonians decouple. For the phase difference we obtain the sine--Gordon Hamiltonian~\cite{Gritsev2007a}
\be
H_-=\frac{\hbar c}2 \int\ud x \left[\frac{2\pi}{K}  \Pi_-(x)^2+\frac{K}{2\pi} [\p_x\fii_-(x)]^2\right]- \int\ud x\,2\Delta\cos[\fii_-(x)]\,.
\ee

\section{Fluctuations of the phase}

We are interested in the fluctations of $\fii_-$ around the minima of the cosine potential. We will calculate this in the harmonic approximation, i.e. by expanding the cosine up to quadratic order which yields
\be
\label{H3}
H_-\approx\frac{\hbar c}2 \int\ud x \left[\frac{2\pi}{K}  \Pi_-(x)^2+\frac{K}{2\pi} [\p_x\fii_-(x)]^2+\frac{2\Delta}{\hbar c} \fii_-(x)^2\right]\,,
\ee
a free massive boson theory with mass gap
\be
m_0=\sqrt{\frac{4\Delta\hbar\pi}{c^3K}}
=\sqrt{\frac{4\Delta m}{nc^2}}\,.
\ee
%
%
The mode expansion of the fields are
\begin{align}
\fii(x) &= \sqrt{\frac{\pi c}{K}}\int\frac{\ud p}{2\pi} \frac1{\sqrt{\w(p)}}\left[b_pe^{ipx/\hbar}+b_p^\dag e^{-ipx/\hbar}\right]\,,\\
\Pi(x) &= -\frac{i}{2\hbar} \sqrt{\frac{K}{\pi c}}\int\frac{\ud p}{2\pi}\sqrt{\w(p)}\left[b_pe^{ipx/\hbar}-b_p^\dag e^{-ipx/\hbar}\right]
\end{align}
with $\w(p)=\sqrt{p^2c^2+m_0^2c^4}$ and $[b_p,b_p^\dag]=2\pi\delta(p-p').$

The finite temperature equal time correlation function of the phase is
\be
\vev{\fii(x)\fii(x')}_T=\frac{\pi c}{K}\int\frac{\ud p}{2\pi}\frac1{\w(p)}e^{ip(x-x')/\hbar}\coth(\beta\w(p)/2)\,,
\ee
where we used $\vev{b^\dag_p b_{p'}}=2\pi\delta(p-p')f_T(p)$
with $f_T(p)=1/(e^{\w(p)/(\kB T)}-1)$ the thermal Bose--Einstein distribution and $\beta=1/(\kB T).$

\subsection{Quantum fluctuations ($T=0$)}

At zero temperature the integral can be evaluated in closed form,
\be
\vev{\fii(x)\fii(x')}_{T=0} = \frac{\pi c}{K}\int\frac{\ud p}{2\pi}\frac1{\w(p)}e^{ip(x-x')/\hbar} = \frac1{K}K_0[m_0c|x-x'|/\hbar]=\frac1{K}K_0\left(\frac{\Delta x}{l_\Delta}\right)\,,
\ee
where $\Delta x =|x-x'|,$ $K_0(z)$ is the modified Bessel function of the second kind and 
\be
\label{compton}
l_\Delta\equiv\frac{\hbar}{m_0 c} = \sqrt{\frac{\hbar^2n}{4m\Delta}} 
\ee
is the Compton wavelength associated with the mass $m_0$ which physically corresponds to the ``healing length of the relative phase''. For small separation 
\be
\vev{\fii(x)\fii(x')}_{T=0} = -\frac1{K} \left\{\log\left(\frac{\Delta x}{2l_\Delta}\right)\left[1+\left(\frac{\Delta x}{2l_\Delta}\right)^2\right] +\gamma_\text{E}\right\}+\mathcal{O}\left[\left(\frac{\Delta x}{l_\Delta}\right)^2\right]\,,
\ee
so it is logarithmically divergent ($\gamma_\text{E}$ is the Euler--Mascheroni constant). By introducing a minimum distance $\alpha$ we obtain
\be
\vev{\fii^2}_{T=0} \approx -\frac1{K} \left[\log\left(\frac{\alpha}{2l_\Delta}\right)+\gamma_\text{E}\right]\,.
\ee
For e.g. $l_\Delta/\alpha \approx 10,$ $\vev{\fii^2}_{T=0} \approx 2.4/K.$ As $K>1,$ this means that the quantum fluctuation of the phase satisfies $\Delta\fii\lesssim1.5$ and remains small compared to $2\pi,$ so the approximation of the cosine by a quadratic potential is justified.

\subsection{Thermal fluctuations (low temperature)}
\begin{figure}[!t]
\includegraphics[width=0.5\columnwidth]{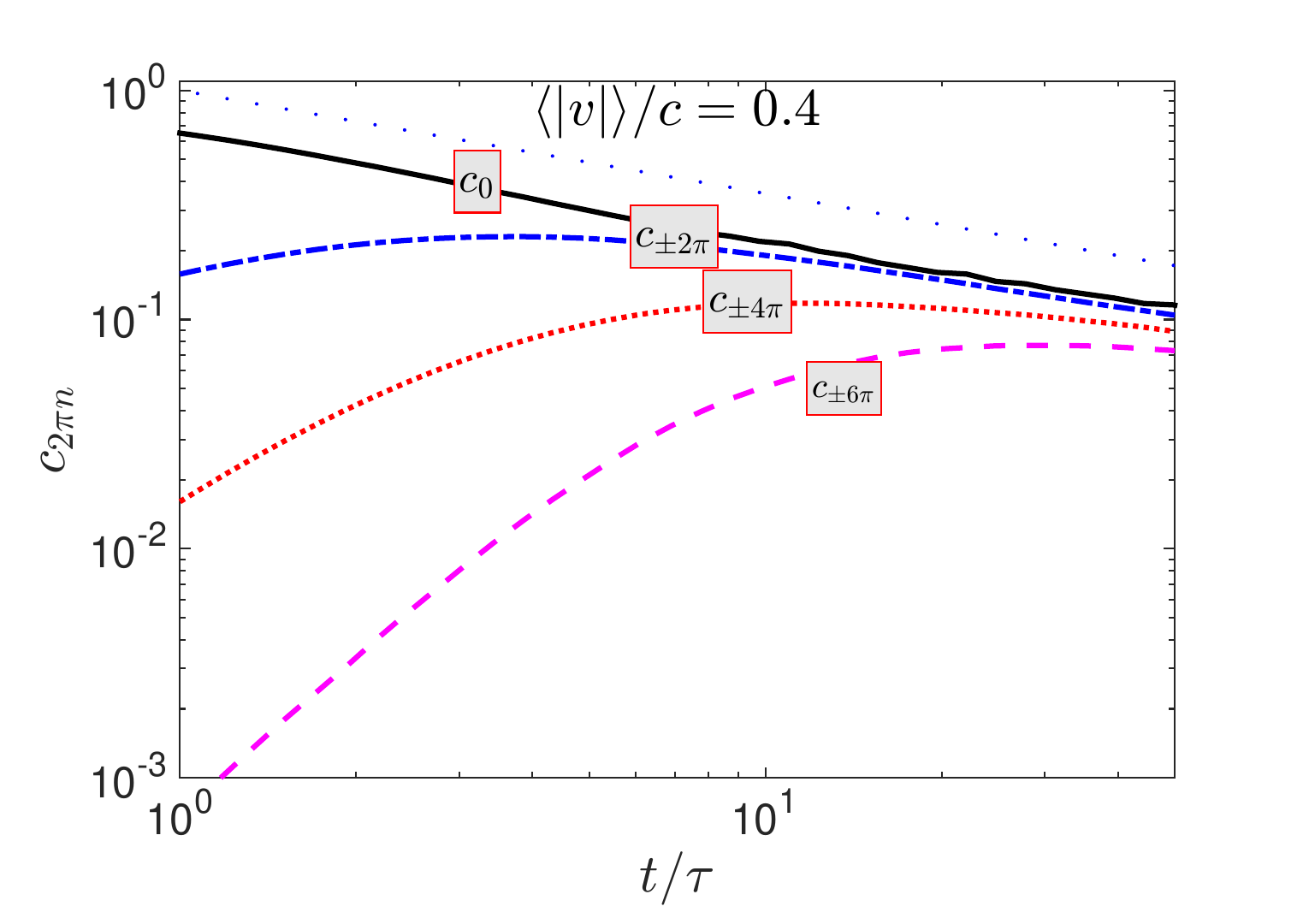}
\caption{(Color online) Time dependence of the  weights $c_{2\pi n}$ defined in Eq.~(5) in the main text
represented on a logarithmic scale. The dotted blue line represents the expected $\sim \sqrt{\tau/t}$
dependence corresponding to diffusive behavior in the large time limit.}
\label{fig:coeffs}
\end{figure}

Let us now compute the correlation function focusing on the low temperature limit, i.e. when $\kB T\ll m_0 c^2.$
The thermal contribution is given by the expression
\be
\vev{\fii(x)\fii(x')}_\text{therm}=\frac{\pi c}{K}\int\frac{\ud p}{2\pi}\frac1{\w(p)}e^{ip(x-x')/\hbar}(\coth(\beta\w(p)/2)-1)\, ,
\ee
which is perfectly well-behaved for large $p.$ Expanding the hyperbolic cotangent in powers of $e^{-\beta\w(p)},$ changing the momentum integration variable to relativistic rapidity, and shifting the integration contour we arrive at
\be
\vev{\fii(x)\fii(x')}_\text{therm}=\frac2{K} \sum_{n=1}^\infty  K_0\left(\sqrt{(n\pi q/2K)^2+(\Delta x/l_\Delta)^2}\right)\,,
\ee
where 
\be
\label{qdef}
q\equiv\frac{\lambda_T}{l_\Delta} =\frac{2K}\pi \frac{m_0c^2}{\kB T}
\ee
is the ratio of the thermal phase coherence length $\lambda_T = 2\hbar^2 n/(m\kB T)$ and $l_\Delta.$
For the fluctuations of $\fii$ this implies
\be
\label{fifluct}
\vev{\fii^2}_T=\vev{\fii^2}_{T=0}+\vev{\fii^2}_\text{therm}
\approx \frac1{K} \left[-\log\left(\frac{\alpha}{2l_\Delta}\right)-\gamma_\text{E}+
 2\sum_{n=1}^\infty  K_0\left(n\frac{\pi q}{2K}\right)\right]\,.
\ee
In the low temperature limit $\pi q/(2K)\gg1$ so we can expand the Bessel functions:
\be
K_0\left(n\frac{\pi q}{2K}\right)=\sqrt{\frac{K}{nq}}e^{-n\pi q/(2K)}\left(1-\frac{K}{4\pi nq}+\mathcal{O}\left[\left(\frac{K}{\pi nq}\right)^2\right]\right)\,.
\ee
Each term in the sum in Eq. \erf{fifluct} is exponentially suppressed with respect to the previous one, so we can truncate the series at the first term, leading to
\be
\label{result}
\vev{\fii^2}_T\approx 
\frac1{K} \left(-\log\left(\frac\alpha{2l_\Delta}\right)-\gamma_\text{E}+2\sqrt{\frac{K}q}e^{-m_0c^2/(\kB T)}\right)\,.
\ee
The calculation is consistent if $\vev{\fii^2}_T\ll1$ so the approximation of the cosine by a quadratic potential is justified.
With the the results for the fluctuations at hand we can dress up the phase distribution (Eq.(5) in the 
main text) by broadening the $\delta$-peaks accordingly. For completeness, we also represent in 
Fig.~\ref{fig:coeffs} the  coefficients  $c_{2\pi n}$ on a log-log scale which shows a decay  $\sim 1/\sqrt{(t/\tau)}$ further supporting the diffusive behavior for the propagation of the phase, as discussed in the main body of the paper. 

\section{S-matrix of the sine--Gordon model}

The 2-particle S-matrix describing the scattering of two kinks is exactly known~\cite{Zamolodchikov1979}. The energy and momentum of the incoming and outgoing particles are conveniently parameterized in terms of the relativistic rapidity as $E=mc^2\cosh\th,$ $p=mc\sinh\th.$ The 2-particle S-matrix is given by a four by four matrix in the basis $|++\rangle,|+-\rangle,|-+\rangle,|--\rangle:$
\be
S=
\begin{pmatrix}
S&&&\\
&S_\text{T}&S_\text{R}&\\
&S_\text{R}&S_\text{T}&\\
&&&S
\end{pmatrix}\,,
\ee
where due to relativistic invariance all entries depend only on the relative rapidity $\theta=\theta_1-\theta_2$ of the two incoming kinks. Here $S_\text{T}$ is the amplitude of transmission and $S_\text{R}$ is the amplitude of reflection. The term $S$ accounts for the phase picked up by the wave function upon scattering of two kinks of the same charge:
\be
S(\theta)=-\exp\left\{-i\int\frac{\ud t}t \frac{\sinh\frac{t(\pi-\xi)}2}{\sinh\frac{\xi t}2\cosh\frac{\pi t}2}\sin(\theta t)\right\}\,,
\ee
where 
\be
\xi=\frac{\pi}{4K-1}\,.
\ee
The transmission and reflection factors are given by
\begin{align}
S_\text{T} (\th)&= \frac{\sinh\frac{\pi\theta}\xi}{\sinh\frac{\pi(i\pi-\theta)}\xi}S(\theta)\,,\\
S_\text{R} (\th)&= i\frac{\sin\frac{\pi^2}\xi}{\sinh\frac{\pi(i\pi-\theta)}\xi}S(\theta)\,.
\end{align}
and they satisfy $|S_\text{T}|^2+|S_\text{R}|^2=1.$ For small $\theta$ they behave as $|S_\text{T}|^2\propto \theta^2,|S_\text{R}|^2\propto1-\theta^2,$ thus at small rapidities the scattering of kinks of opposite charges is almost purely reflective.

\section{Details of the numerical algorithm}

\begin{figure}[!h]
\includegraphics[width=1\columnwidth]{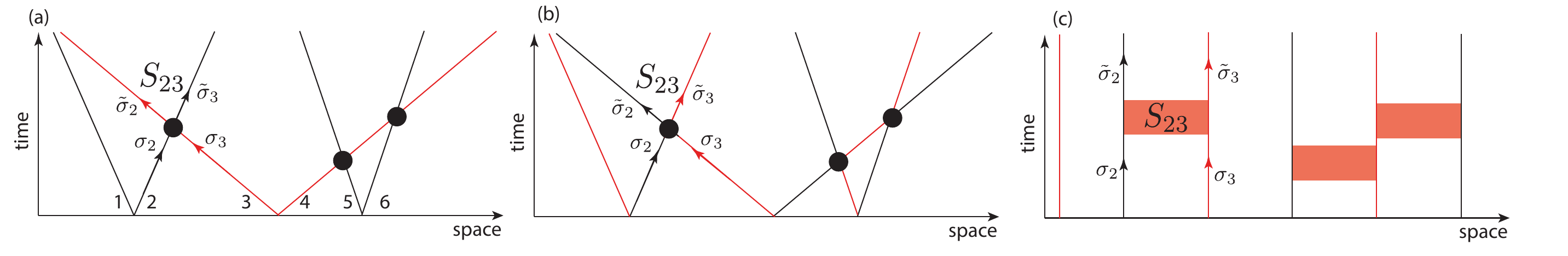}
\caption{(Color online) (a) A typical kink configuration with soliton-antisoliton pairs. Their
space-time evolution is described by pairs of straight lines. Different lines are plotted with 
different colors.  (b) Reindexing the lines after each collision. The quasiparticle trajectories have now zig-zagged shapes indicated by the colors. With this labeling
the collisions always take place between neighboring quasiparticles. 
(c) Propagation of the TEBD solution in the charge sector of the wave function. 
When two lines intersect in (b), the effective spin chain of the charge sector is acted on by the unitary evolution operator given by the corresponding S-matrix. }
\label{fig:sketch}
\end{figure}

In this section we discuss in more detail the numerical algorithm which is a combination 
of Monte Carlo (MC) sampling of classical trajectories for the soliton-antisoliton pairs and
a propagation in time of the initial wave function $\ket{ \Psi_{\rm orb}({\bf x},t)}$
using the Time-Evolving Block Decimation (TEBD) approach. The process can be divided into three main steps as follows:

(1) \emph{Generation of classical kink configurations}-- For a given concentration of pairs, 
we randomly generate  their positions at $t=0$. The kink velocities are generated from a given velocity 
distribution~\footnote{In the present calculation the velocity distribution is derived from a
distribution with relativistic dispersion, but we have checked that the results remain
largely unaltered by using different distributions.}.  Then we construct the 
kink configuration (one such configuration is 
presented in Fig.~\ref{fig:sketch}(a)) which is a space-time diagram in $(x,t)$ coordinates
that displays the classical trajectories of the kinks.
We have to keep in mind that at $t=0$ two kinks that form a pair start to move in 
opposite directions with equal velocities.  In this representation the 
collision of two kinks corresponds to the intersection of two lines. When constructing the 
kink configuration we index all the intersection points in the 
space-time coordinates $(x_{I}, t_{I})$ as well as the corresponding lines. These coordinates are then
ordered chronologically. In this way we completely characterize  the orbital motion of the kinks, which 
corresponds to the construction or the orbital part of the wave function
$\ket{ \Psi_{\rm orb}({\bf x},t)}$ in Eq.~(2b). Furthermore 
(although not displayed in Fig.~\ref{fig:sketch}(a)) we impose hard wall boundary conditions such that 
a kink is perfectly reflected off the walls.

(2) \emph{Construction of the effective spin model}-- Quantum effects become
relevant only at collision times and they do not affect the orbital motion of the quasiparticles, which allows us to factorize the wave function as in Eq.~(2) in the main body of the paper.  
To be able to propagate the initial wave function we need to map the dynamics of the spins to an 
effective spin model. We do that by constructing first zig-zag kink configuration, which 
consists of relabeling the lines according to Fig.~\ref{fig:sketch}(b). In this way, the lines 
are ordered from left to right at any time instance $t$ but, more importantly,  in this 
representation only neighboring lines  intersect. 
Furthermore, each line $j$ carries a ``spin'' $\sigma_j$. The intersections of classical lines corresponds to a scattering event of the two spins carried by the lines. This is 
a quantum mechanically process that is fully described in terms of the S-matrix, so 
at this point the model Hamiltonian is not relevant 
as all the necessary information that we need is encoded in the two-body S-matrix. 

(3) \emph{Time evolution of the wave function}-- The mapping of the zig-zag kink configuration 
to the effective spin chain model is displayed in Fig.~\ref{fig:sketch}(c). Here the square regions
indicate the interaction between two neighboring spins. 
Basically, the spin chain is frozen in time  in between collisions and 
the time evolution takes place 
only when pairs of spins are scattered at the collision times. 
This picture allows us to use the TEBD algorithm~\cite{Vidal2003, Schollwock:2011}
and propagate the initial wave function. 
The $t=0$ wave function $\ket{\chi(\boldsymbol \sigma, t=0)}$ given in Eq.~(3)
is first organised as an MPS state which is then evolved in time. The full time evolution
operator $U(t)$ is the chronological product of unitary two-body S-matrix operators that scatter
pairs of quasiparticles only at the intersection times $t_{I}.$ Furthermore, the TEBD framework
allows us to compute different physical quantities~\cite{Vidal2003}. For example,  to compute the 
time evolution of the entanglement entropy, we first evolve the MPS state up to the desired time $t,$
bipartition the state using the Schmidt decomposition and then compute the 
entanglement entropy in the usual way~\cite{Vidal2003}.

What we have described so far is the evolution of a single kink configuration. To compute the averages
discussed in the main body of the paper, we sample in general $\sim 10^5$ such configurations and
average over their positions and velocity distribution.

\end{document}